\documentclass[
preprint,
 amsmath,amssymb,
 aps,
]{revtex4-1}

\usepackage{graphicx}
\usepackage{dcolumn}
\usepackage{bm}
\usepackage{hyperref}

\begin{document}

\preprint{APS/123-QED}

\title{A Deep Dive into $f(R)$ Gravity Theory}

\author{Solmaz Asgari}
\email{asgari\underline{ }s@iau$-$abhar.ac.ir}
 \affiliation{Department of Science,
Islamic Azad University, Abhar Branch, P.O.Box 22, Abhar, Iran.}
\author{Reza Saffari}%
 \email{rsk@guilan.ac.ir}
\affiliation{Department of Physics, University of Guilan, P.O.Box
41335-1914, Rasht, Iran. }

\date{\today}

\begin{abstract}
In this paper we have derived the behavior of deceleration parameter
with respect to redshift in context of $f(R)$ gravity in vacuum
using Taylor expansion of derivative of action. Here we have
obtained that the two first terms in Taylor expansion may describe
the late time acceleration which is appeared by SNeIa without need
of dark energy and dark matter. Also we have derived that any other
terms higher than $z$ in Taylor expansion may describe main
inflationary epoch in the early Universe. We have shown that $f(R)$
gravity may cover all the dynamical history of the Universe from the
beginning to the late time accelerating phase transition.
\end{abstract}

\pacs{Valid PACS appear here}
\maketitle



\begin{figure}
\includegraphics{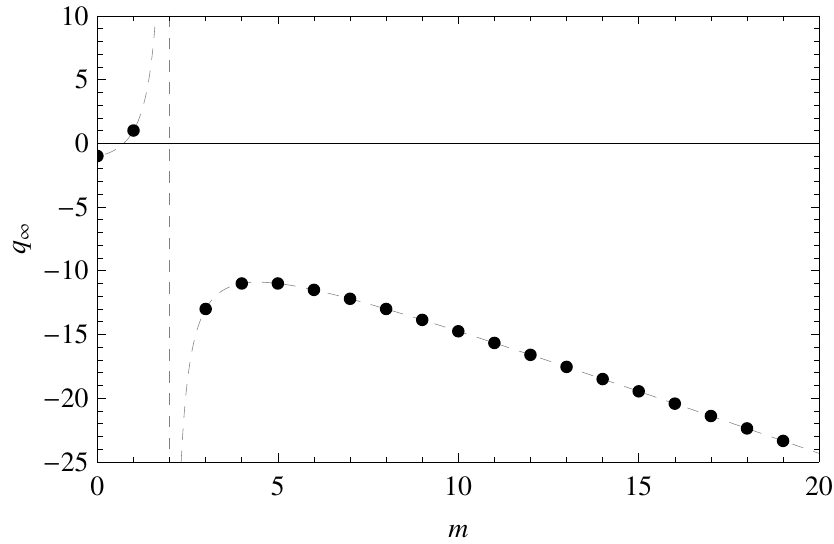}
\caption{\label{numsen}Evolution of deceleration parameter in
$z\rightarrow\infty$ with respect to the number of sentences in the
Taylor expansion of $F(z)$.}
\end{figure}

\begin{figure}
\includegraphics{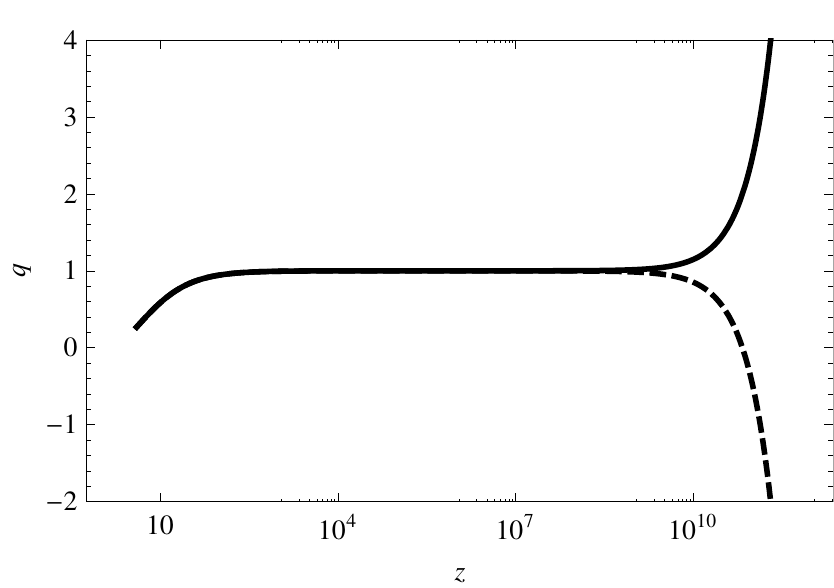}
\caption{\label{z2}Evolution of deceleration parameter which is
introduced in Eq. (\ref{simtey}) for $\gamma=0$, $\beta>0$ solid
line and $\beta<0$ dashed line.}
\end{figure}

\begin{figure}
\includegraphics{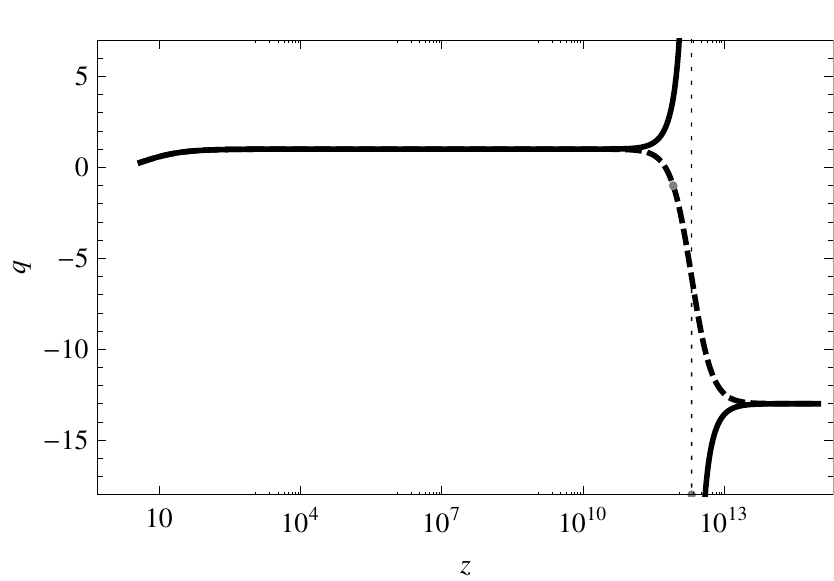}
\caption{\label{z3}Evolution of deceleration parameter which is
introduced in Eq. (\ref{simtey}) for $\beta=0$, $\gamma>0$ solid
line and $\gamma<0$ dashed line.}
\end{figure}

\begin{figure}
\includegraphics{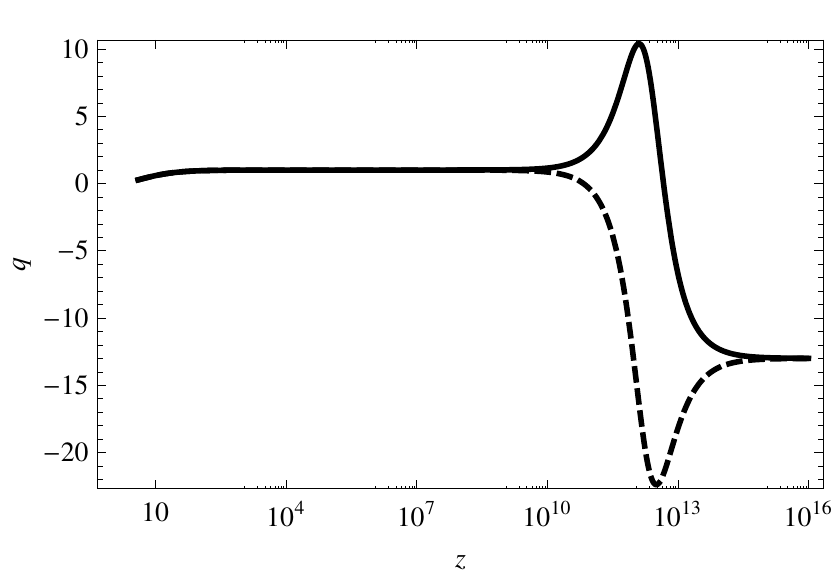}
\caption{\label{z23}Evolution of deceleration parameter which is
introduced in Eq. (\ref{simtey}) for the case $\beta>0$ and
$\gamma<0$ solid line and the case $\beta<0$ and $\gamma<0$ dashed
line.}
\end{figure}

\begin{figure}
\includegraphics{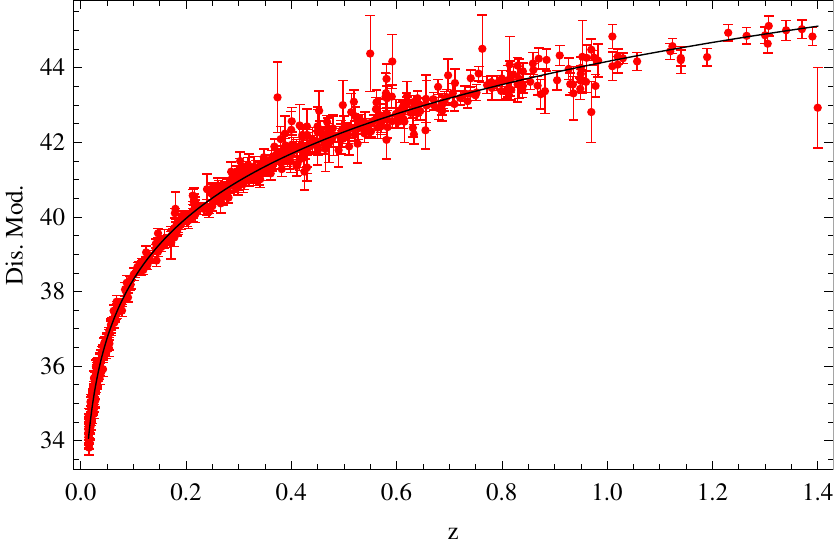}
\caption{\label{dismod}Accordance between Union2 SNeIa data and late
time approximation of Eq. (\ref{simtey}) for $\alpha=0.41$ and
$h=0.7$.}
\end{figure}

\begin{figure}
\includegraphics{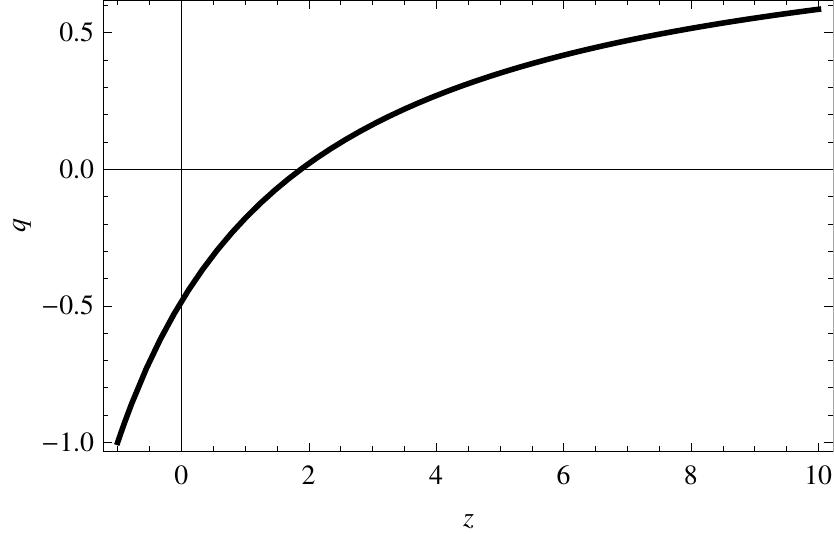}
\caption{\label{latetime}Late time evolution of deceleration
parameter for $\alpha=0.41$ and $h=0.7$.}
\end{figure}

The recent data coming from the luminosity distance of SuperNovae Ia
(SNeIa)\cite{riess}, wide galaxy surveys \cite{cole} and the
anisotropy of cosmic microwave background radiation
\cite{sper03,sper07} suggest that the Universe is undergoing an
accelerating expansion. Several candidates, responsible for this
expansion, have been proposed in the literature, in particular, dark
energy models and modified gravity. The main problem of dark energy
models are understanding their nature, since they are introduced as
ad hoc gravity sources in a well define models of gravity. In this
context, modified theories of gravity, such as f(R) gravity
\cite{Nojiri7,Capozziello,Faraoni}, has been verified in an attempt
to explain the late-time accelerated expansion of the Universe.
These theories are also referred to as 'extended theories of
gravity', since they naturally generalize General Relativity. There
has been predicted that the universe might have appeared from an
inflationary phase in the past. It is also believed that the present
universe is passing through a phase of the cosmic acceleration.
While a series of cosmological models were proposed in Einstein's
gravity with early inflationary scenario are working well, the
mystery of passing universe through an accelerated phase of
expansion is an interesting open question that is yet to be
understood. There are many interesting works have done in the
context of $f(R)$ gravity on unifying late time accelerating
behavior and early inflationary behavior of the Universe
\cite{Nojiri}, where they have shown that the late time acceleration
is in continues of early time inflation for specific models of
$f(R)$ gravity. Also, their vacuum solutions are obtained for
constant Ricci scalar while it is possible to derive non constant
curvature scalar vacuum solutions in $f(R)$ gravity theories. In
this letter we show easily that $f(R)$ gravity theories are the most
powerful approach to describe all the dynamical behavior of the
Universe in one scenario. Here we look to the story of $f(R)$
gravity from a different angle. Actually we consider vacuum
solutions of $f(R)$ gravity. But there are two main difference with
other vacuum solutions. The first one is that we do not assume
constant scalar curvature to obtain vacuum solution. The second one
is that we do not consider any specific model of $f(R)$ gravity
action. Vacuum solutions of modified $f(R)$ gravity would like to
explain all the phase transition of cosmological parameters like
deceleration parameter without need of dark companion of the
Universe, alone by pure geometry.
The action of modified theory of gravity is given by
\begin{equation}
S=\int d^4x\sqrt{-g}
\bigg[\frac{1}{2\kappa}f(R)+L_m\bigg],\label{action}
\end{equation}
where $L_m$ is the matter action such as radiation, baryonic matter,
dark matter and so on which we do not consider in field equation. In
this work, we consider the flat Friedmann Robertson Walker, (FRW)
background, so that the gravitational field equations for $f(R)$
modified gravity are provided by the following form
\begin{equation}
-3\frac{\ddot{a}}{a}f'+3\frac{\dot{a}}{a}\dot{R}f''
+\frac12f=0,\label{modfrw1}
\end{equation}
\begin{equation}
[\frac{\ddot{a}}{a}+2\frac{\dot{a}^2}{a^2}]f'
-2\frac{\dot{a}}{a}\dot{R}f''-\dot{R}^2f'''
-\ddot{R}f''-\frac12f=0.\label{modfrw2}
\end{equation}
where the overdot denotes a derivative with respect $t$, $a(t)$ is
the scale factor and $H=\dot{a}(t)/a(t)$ is the Hubble parameter.
Eliminating $f$ between Eqs. (\ref{modfrw1}) and (\ref{modfrw2})
obtains:
\begin{equation}
-2[\frac{\ddot a}{a}-(\frac{\dot a}{a})^2]f'+\frac{\dot a }{a}\dot
Rf''-\ddot Rf''-\dot R^2f'''=0 .\label{elim}
\end{equation}
Which can be changed to the form of:
\begin{equation}
\ddot F-H\dot F+2\dot HF=0,\label{diff1}
\end{equation}
where $F=df/dR$. Eq. (\ref{diff1}) is a second order differential
equation of $F$ with respect to time, in which both of $F$ and $H$
are undefined. The usual method to solve Eq. (\ref{diff1}) is based
on definition of $f(R)$. But in this paper we would like to replace
the variable of Eq. (\ref{diff1}) by another cosmic parameter which
named redshift, $z$. Each redshift, $z$ has an associated cosmic
time $t$ (the time when objects observed with redshift $z$ emitted
their light), so we can replace all the differentials with respect
to $t$ by $z$ via:
\begin{eqnarray}
\frac{d}{dt}&=&\frac{da}{dt}\frac{dz}{da}\frac{d}{dz},\nonumber\\
&=&-(1+z)H(z)\frac{d}{dz},\label{rep1}
\end{eqnarray}
where we use $1+z={a_0}/{a}$, and we consider $a_0=1$, in the
present time. Now, we can replace the variable of Eq. (\ref{diff1})
from $t$ to $z$ by using Eq. (\ref{rep1}), and we obtain a first
order differential equation for $H^2$ with respect to $z$ as:
\begin{equation}
\frac{d}{dz}H(z)^2=P(z)H(z)^2,\label{diff3}
\end{equation}
where $P(z)$ is a function with respect to $z$, which depend on
definition of $F$ with respect to $z$ as:
\begin{equation}
P(z)=\frac{2(1+z)(d^2F/dz^2)+4(dF/dz)}{2F-(1+z)(dF/dz)}.\label{pz}
\end{equation}
The deceleration parameter, $q$ in cosmology is a dimensionless
measure of the cosmic acceleration of the expansion of space in a
FRW universe. It is defined by:
\begin{equation}
q=-\frac{\ddot aa}{\dot a^2}=-1-\frac{\dot H}{H^2}.\label{q}
\end{equation}
here we change the variable of Eq. (\ref{q}) from $t$ to $z$ to have
evolution of deceleration parameter with respect to redshift as:
\begin{eqnarray}
q(z)&=&-1+\frac{1+z}{2H(z)^2}\frac{dH(z)^2}{dz},\label{qz1}\\
&=&-1+\frac{1+z}{2}P(z),\label{qz2}
\end{eqnarray}
which only depends on $P(z)$ for vacuum solution. Since we would
like to consider a general behavior of $q(z)$ due to the model
independent $F(z)$, we use Taylor expansion of $F(z)$ around the
present value of redshift, $z=0$ as:
\begin{eqnarray}
F(z)=\sum_{n=0}^{m} F_nz^n\label{teyf}
\end{eqnarray}
where
\begin{equation}
F_n=\frac{1}{n!}\frac{d^nF}{dz^n}\bigg|_{z=0},\label{abg}
\end{equation}
are some constants at $z=0$. Now to calculate deceleration
parameter, with some algebra we put Taylor expansion of Eq.
(\ref{teyf}) in Eq. (\ref{pz}), then we put the result in Eq.
(\ref{qz2}) to find:
\begin{equation}
q(z)=-1+\frac{(1+z)\displaystyle\sum_{n=0}^{m}nF_nz^{n-2}[(n+1)z+n-1]}
{\displaystyle\sum_{n=0}^{m}F_nz^{n-1}[(2-n)z-n]},\label{qtey}
\end{equation}
which is the model independent deceleration parameter in a redshift
based $f(R)$ gravity.
The present time deceleration parameter for a general $F(z)$ obtains
as:
\begin{equation}
q_0\equiv q(z=0)=-1+\frac{\alpha+\beta}{1-\alpha/2},\label{q0}
\end{equation}
where $\alpha=F_1/F_0$, $\beta=F_2/F_0$ which shows that the present
value of deceleration parameter depends on, only the first two terms
coefficients of Taylor expansion of $F(z)$. Since the present value
of deceleration parameter should be negative, $q_0<0$, $\alpha$ and
$\beta$ will constrained by $3\alpha/2<1-\beta$.
In the early time, $z\rightarrow\infty$ the only remained terms in
the numerator and denominator of the right hand side of Eq.
(\ref{qtey}) are the terms with higher powers in $z$ as:
\begin{equation}
q_\infty(m)\equiv q(z\rightarrow\infty)=
-1+\frac{m(m+1)}{2-m},\label{qinf}
\end{equation}
which is independent of coefficients of Taylor expansion of $F(z)$.
Here we see that $q_\infty$ depends on the number of the sentences
in Taylor expansion. For example if $m=0$ (one term in Taylor
expansion), we will have a constant modification, which will recover
de' Sitter space, $q_\infty(0)=-1$. If $m=1$ (two terms in Taylor
expansion), the value of deceleration parameter is $q_\infty(1)=+1$,
which is consistent with decelerating Universe after inflationary
epoch. If $m=2$ (three terms in Taylor expansion), we have
$q_\infty(2)\rightarrow\pm\infty$, which is an unstable value for
deceleration parameter. Eventually for any values of $m\geqslant3$,
the early time value of deceleration parameter is a negative number,
$q_\infty(m\geqslant 3)<-1$, which shows a continues inflationary
behavior. One can show that adding any extra term higher than first
order term in Taylor expansion can make an unstable epoch in
evolution of deceleration parameter. But location of this
instability which we may call it, inflation, depends on the value of
coefficients of Taylor expansion. Dependency of $q_\infty$ on the
number of sentences in Taylor expansion of $F(z)$ is plotted in Fig.
(\ref{numsen}).
As a very simple but phenomenological case we can assume there are
four nonzero terms in Taylor expansion as:
\begin{eqnarray}
F(z)=F_0(1+\alpha z+\beta z^2+\gamma z^3),\label{simtey}
\end{eqnarray}
where $\gamma=F_3/F_0$. Depend on choosing expansion coefficients we
can see evolutionary behavior of deceleration parameter with respect
to redshift. There has plotted three different cases for early time
period in Figs. (\ref{z2}-\ref{z23}). The common result of three
above figures is that, There is a long era between early time and
late time evolution of deceleration parameter in which $q(z)$ is
equal to $+1$ in each figure. This period is called radiation
dominated era. Before that in higher $z$'s there are different
instabilities. The other common point is that they desire to have an
acceleration phase transition point in late time Universe.
In Fig. (\ref{z2}) $\beta=\pm 10^{-12}$ and $\gamma=0$. For the
positive sign (solid line) of $\beta$ the Universe may begin from an
infinite decelerating state which is continued to the end of
radiation dominated era. For the negative case (dashed line) the
Universe may begin from an infinite accelerating state which is
continued to the beginning of radiation dominated era. There is one
acceleration transition point in early time in this latter case. In
Fig. (\ref{z3}) we choose $\beta=0$ and $\gamma=\pm 10^{-25}$, then
for the positive sign (solid line) the Universe may begin from a
finite accelerating state. But it is faced to an instability in a
identified redshift which is called inflation. Therefor deceleration
parameter goes to $-\infty$ and then comes back to a finite value
from $+\infty$. For the negative case (dashed line) the Universe
again may begin from the same infinite accelerating state which goes
to the transition point in a continuous state up to join to
beginning of the radiation era. In Fig (\ref{z23}) there are two
cases of composition of values of coefficients. In the first one
$\beta=+10^{-12}$ and $\gamma=-10^{-25}$ (solid line) which shows
that the Universe may begin from a finite accelerating state and
then goes to a finite maximum deceleration value and then comes back
to radiation era. There could not be an inflationary period in this
case. While in the other case $\beta=-10^{-12}$ and
$\gamma=-10^{-25}$ (dashed line) there could be an stable inflation.
Here we saw that smallness of the coefficients $\beta$ and $\gamma$
may move any inflationary behavior to early Universe. However this
example of composition of expansion of coefficients are nor
realistic but will make this sense that we can find a close
combination which might be compatible with number of e-folding and
other early time cosmological constraints. The other main late time
cosmological constraint which is tested for the above simple example
is SNeIa distance module with respect to redshift constraint. In
this comparison we used the Union2 data set \cite{union2} which is
provide 557 SNeIa specifications.
According to the value of numerical results which was taken in
comparison with SNeIa data set from usual $\chi^2$ algorithm, we
obtain that for $\alpha=0.41$, $h=0.7$ and no need to dark
companions of the Universe, $\chi^2=546.48$ and the ratio of chi
square error to the number of freedom is about $0.98$. The result of
this calculation is shown in Fig. (\ref{dismod}). According to the
obtained value for of parameter $\alpha$, evolution of deceleration
parameter for low redshift Universe is plotted in Fig.
(\ref{latetime}). Other cosmological parameters such as jerk, snap
and lerk is calculated for this combination of coefficients. The
obtained values of present cosmological parameters $(q_0, j_0, s_0,
l_0)$ are $(-0.48, 0.37, -0.03, 0.71)$ while their values at the
beginning of radiation era $(q_r, j_r, s_r, l_r)$ obtain as $(1, 3,
-15, 105)$. The same values of present cosmological parameters from
$\Lambda CDM$ for $\Omega_M=0.3$ and $h=0.72$ is $(-0.55, 1.0,
-0.35, 3.11)$.
In this letter we have considered a general effect of changing the
gravitational action on cosmological deceleration parameter, from
$R$ to $f(R)$. Here we have obtained that all the cosmic history may
describe by knowing effect of any sentences in Taylor expansion of a
generic function $F(z)$. The first term which is a constant will
recover a de' Sitter Universe, while the first two terms have cover
all dynamical properties of the Universe from radiation era to late
accelerating Universe. No one extra term will affect on the present
value of deceleration parameter. Adding the third term to the last
two terms may make an inflationary behavior, in which deceleration
parameter may goes back to a huge positive or negative amount.
Adding any other terms to the last three terms may change
the value of deceleration parameter from an unstable value to a
stable but negative value in very early Universe. It means that in
$f(R)$ theory of gravity in the absence of matter, the value of
deceleration parameter was not a positive value which denotes there
was a low inflationary era before main inflation. Adding more terms
to the last four terms do not change whole the story, but will vary
the value of deceleration parameter to a larger value in negative
direction. Thus for an infinite number of terms in Taylor expansion
of $F(z)$ the value of deceleration parameter will goes to
$-\infty$. It means the Universe is born in a main inflationary
state. Then if the number of Taylor expansion sentences should be
finite whether the Universe was born in an inflationary state.
However the value of expansion coefficients may make some
fluctuations between the first moment of creation and the beginning
of radiation era in $f(R)$ gravity theory.



\end{document}